\documentclass[%
 reprint,
superscriptaddress,
 amsmath,amssymb,
 aps,
]{revtex4-1}
\usepackage{adjustbox}
\usepackage{subfig}
\usepackage{graphicx}
\usepackage{dcolumn}
\usepackage{bm}

\makeatletter
\newcommand{\thickhline}{%
    \noalign {\ifnum 0=`}\fi \hrule height 1pt
    \futurelet \reserved@a \@xhline
}
\newcolumntype{"}{@{\hskip\tabcolsep\vrule width 1pt\hskip\tabcolsep}}
\makeatother

\begin{document}

\preprint{APS/123-QED}
\graphicspath{ {Images/}}
\title{Splashing of impacting drops}

\author{T. C. de Goede}
\email{T.CdeGoede@uva.nl}
\affiliation{ Van der Waals-Zeeman Institute, Institute of Physics, University of Amsterdam, Science Park 904, 1098 XH Amsterdam, Netherlands\\
} 
\author{K.G. de Bruin}
\affiliation{ Van der Waals-Zeeman Institute, Institute of Physics, University of Amsterdam, Science Park 904, 1098 XH Amsterdam, Netherlands\\
}
\affiliation{Netherlands Forensic Institute, Laan van Ypenburg 6, 2497 GB The Hague, Netherlands\\ }

\author{D. Bonn}
\email{D.bonn@uva.nl}
\affiliation{
 Van der Waals-Zeeman Institute, Institute of Physics, University of Amsterdam, Science Park 904, 1098 XH Amsterdam, Netherlands\\
}

\date{\today}

\begin{abstract}
We investigate the impact velocity beyond which the ejection of smaller droplets from the main droplet (splashing) occurs for droplets impacting a smooth surface. We examine its dependence on the surface wetting properties and droplet surface tension. We show that the splashing velocity is independent of the wetting properties of the surface, but increases roughly linearly with increasing surface tension of the liquid. A preexisting splashing model is considered that predicts the splashing velocity by incorporating the air viscosity. The model is consistent with our data, but the calculation is complex. To address this issue, we propose a simplification, assuming atmospheric conditions and low viscosities, and show that the simplification gives an equally good prediction of the splashing velocity from a simple analytical formula.

\end{abstract}

\pacs{Valid PACS appear here}
\maketitle
When impacting a dry, smooth surface, a droplet either spreads over the surface for low impact velocities or disintegrates into smaller droplets for high impact velocities. This so-called splashing phenomenon has been the subject of numerous studies for the last several decades and is of relevance for a wide range of practical applications like crop spraying \cite{Wirth1991,Bergeron2000}, rain drops impacting on porous stones \cite{Shahidzadeh2008,Lee2016} and forensic research \cite{Knock2007,Adam2013,Laan2015}. The splashing velocity $v_{\rm sp}$ is defined as the critical value of the impact velocity of the droplet beyond which splashing occurs. Many studies have tried to find an empirical relation between splashing velocity and fluid parameters \cite{Stow1981,Yarin1995,Mundo1995,Bussmann2000,Wal2006,Stevens2014} and between splashing velocity and surface properties \cite{Stow1977,Wu1992,Range1998,Latka2012}. In addition, Xu \textit{et al.} \cite{Xu2005} showed that the atmospheric conditions have a significant influence on droplet splashing \cite{Stevens2014,Latka2012,Stevens2014b}, implying that the air viscosity is also an important parameter. To include the air viscosity, Riboux and Gordillo recently proposed a theoretical model for impact and splashing on smooth surfaces \cite{Riboux2014}. They postulated that splashing occurs due to the break up of a small liquid film that lifts off the surface just after impact due to the lift force generated by the surrounding air. In the model, the fluid and substrate properties govern the splashing, as well as the air viscosity; in \cite{Riboux2014} quantitative agreement was found between the model and experiments on a single substrate.\\

Numerous studies have investigated splashing \cite{Stow1981,Yarin1995,Mundo1995,Bussmann2000,Wal2006,Stevens2014,Stow1977,Wu1992,Range1998,Latka2012,Xu2005,Stevens2014,Latka2012,Stevens2014b}; however, the influence of the wetting properties of the surface has not been considered in detail. It has recently been shown that the wetting properties of the surface affect droplet spreading at low impact velocities \cite{Laan2014,Lee2015}. In the model of Riboux and Gordillo, the splashing velocity also depends on the surface properties \cite{Riboux2014}. The surface tension and the wetting properties of the surface are closely linked through Young's Law \cite{Bonn2009}. Therefore, the dependence of the splashing velocity on the surface tension should also be taken into account. \\
\begin{table*}[htb]
\centering
\caption{Ethanol mass fraction, density, surface tension and viscosity values of the water, ethanol and ethanol-water mixtures used in this study. Source: \cite{Haynes2014,Vazquez1995}}
\renewcommand{\arraystretch}{2}
\centerline{
  \begin{tabular}{| c | c | c | c |}
  \hline
    wt(\%)& Density $\left( \frac{kg}{m^3} \right)$ & Surface tension  $\left( \frac{mN}{m} \right)$ & Viscosity  $\left( mPa\cdot s \right)$\\
  \hline
   0 & 997.0 & 71.99  & 0.89 \\
  \hline
   5 & 989.0 & 56.41  & 1.228 \\
  \hline
    10 & 981.9 & 48.14  & 1.501 \\
  \hline
  15 & 975.3 & 42.72  & 1.822 \\
  \hline
   20& 968.7 & 37.97 & 2.142 \\
  \hline
   40 & 935.3 & 30.16 & 2.846 \\
  \hline
   60 & 891.1 & 26.23 & 2.547 \\
  \hline
   80 & 843.6 & 23.82 & 1.881 \\
  \hline
   100 &789.3 & 21.82 & 1.203 \\
   \hline
  \end{tabular}}
  \label{tab:fluidparm}
\end{table*} 

In this paper, we systematically investigate the effect of the surface tension of the liquid and the wetting properties of the surface on the splashing velocity of the droplet. Using high-speed camera footage, we measure the splashing velocity of a set of ethanol-water mixtures impacting on different surfaces. We compare our results with the splashing model of Riboux and Gordillo \cite{Riboux2014}. Although the model is consistent with the experimental data, the calculation is complex and depends on several parameters that have to be inferred from the experimental conditions and have to be calculated separately. To address this shortcoming, we propose a simplification valid for low Ohnesorge numbers and atmospheric conditions, which is the situation that pertains to most practical applications and show that it predicts the splashing velocity very well. \\

\begin{figure*}[htb]
    \centering
    \begin{tabular}{cc}
    \adjustbox{valign=b}{\begin{tabular}{@{}c@{}}
    \subfloat[$v < v_{\rm sp}$ \label{fig:vcrita}]{%
          \includegraphics[width=.4\linewidth]{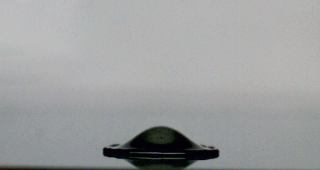}}\\
    \subfloat[$v\geq v_{\rm sp}$ \label{fig:vcritb}]{%
          \includegraphics[width=.4\linewidth]{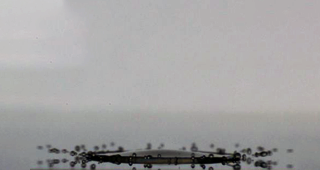}}
    \end{tabular}}
    &
     \adjustbox{valign=b}{\subfloat[\label{fig:vcritc}]{%
          \includegraphics[width=.5\linewidth]{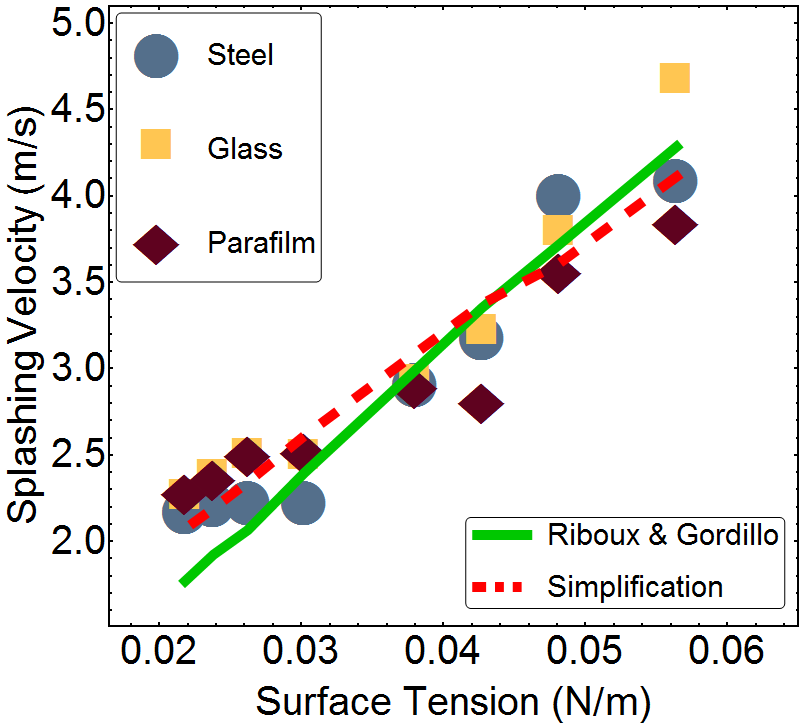}}} 
    \end{tabular}
    \caption{(a) Droplet impact for $v < v_{\rm sp}$ where no splashing occurs. (b) Droplet impact for $v\geq v_{\rm sp}$ where splashing occurs. (c) Measured critical velocity as function of the surface tension for a stainless steel (blue circles) and borosilicate glass (yellow squares). The green line depicts the splashing velocity as given by the full splashing model [Eq.~\eqref{eq:fullmodel}] while the red dashed line shows the splashing velocity as calculated from the simplified model [Eq.~\eqref{eq:critsimpl}].}\label{fig:vcrit}
  \end{figure*}

 In order to measure the splashing velocity $v_{\rm sp}$, droplet impacts were recorded using a high-speed camera (Phantom Miro M310). The droplets were generated from a blunt tipped needle (needle diameter $0.4$ mm) using a syringe pump, where the needle was suspended above the substrate at a certain height. By systematically increasing the height of the needle, and checking whether the droplet merely spreads over the surface (Fig.~\ref{fig:vcrita}) or splashes (Fig.~\ref{fig:vcritb}) for each height, we determined the initial droplet diameter $D_0$ and impact velocity $v$ at the onset of splashing for each liquid. The fluid parameters of each liquid are given in Table~\ref{tab:fluidparm}. Three surfaces were investigated: two hydrophilic surfaces (stainless steel and borosilicate glass) and one hydrophobic surface (parafilm). \\

The measured splashing velocities are plotted as function of surface tension in Fig.~\ref{fig:vcritc}. The graph shows that the splashing velocity increases roughly linearly with the liquid surface tension. For pure water droplets, no splashing was observed within the velocity range investigated here ($0.1 < v < 4.7 $ m/s). No significant difference between the three evaluated surfaces is observed, implying that the splashing velocity is independent of the wetting properties of the substrate.\\

To describe splashing, the air viscosity needs to be considered. To do so, in \cite{Riboux2014} the ejection time $t_e$ (the moment a thin liquid sheet appears from the droplet after impact) is calculated numerically using the momentum balance equation:

\begin{align}
\frac{\sqrt{3}}{2} Re^{-1}t_e^{-\frac{1}{2}} + Re^{-2}Oh^{-2} = 1.21 t_e^\frac{3}{2} 
\label{eq:numte}
\end{align}
 
where $Re = \frac{\rho R v}{\mu}$ and $Oh = \frac{\mu}{\sqrt{\rho R \sigma}}$ are the Reynolds and Ohnesorge numbers, respectively; $\rho$ the density, $\mu$ the viscosity, $\sigma$ the surface tension, $R$ the radius and $v$ the impact velocity of the droplet. Using the ejection time, the velocity $V_t $ and thickness $H_t$ of the thin liquid sheet can be calculated:

\begin{subequations}
\begin{align}
V_t = \frac{1}{2}\sqrt{3}\hspace{0.1cm} v \hspace{0.1cm} t_e^{-\frac{1}{2}}
\label{eq:vtfull}
\end{align}
\begin{align}
H_t = \frac{\sqrt{12}}{\pi}\hspace{0.1cm} R\hspace{0.1cm} t_e^\frac{3}{2}
\label{eq:htfull}
\end{align}
\end{subequations}

\begin{figure}[tb]
\centering
\subfloat[$t= 0.137 \hspace{0.1cm} \text{ms}$]{\label{fig:wedge1}\includegraphics[width=0.24\textwidth]{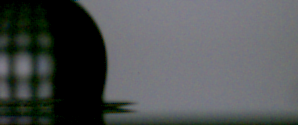}}\hfill
\subfloat[$t= 0.342 \hspace{0.1cm} \text{ms}$]{\label{fig:wedge2}\includegraphics[width=0.24\textwidth]{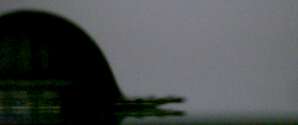}}\\
\subfloat[$t= 0.548 \hspace{0.1cm} \text{ms}$]{\label{fig:wedge3}\includegraphics[width=0.24\textwidth]{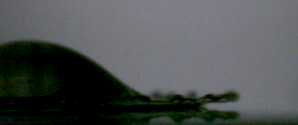}}\hfill
\subfloat[$t =0.685 \hspace{0.1cm} \text{ms}$]{\label{fig:wedge4}\includegraphics[width=0.24\textwidth]{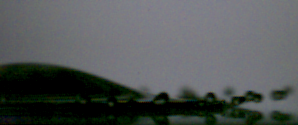}}\\
\subfloat[]{\label{fig:wedgepoints}\includegraphics[width = .45\textwidth]{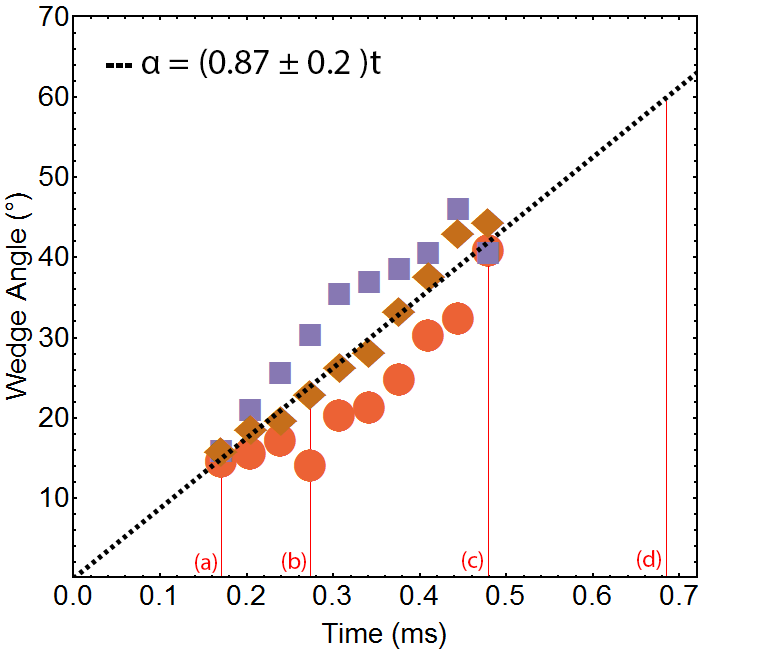}}
\caption{(a)-(d) High-speed camera footage of the lifted liquid sheet after impact (with $t = 0 \hspace{0.1cm} \text{s}$ the moment of droplet impact). (e) Measured wedge angle $\alpha$ as function of the time as deduced from three independent splashing experiments on stainless steel. The black dotted line is the best linear fit of the wedge angle measurements. The vertical red lines depict the different stages of the liquid sheet spreading as given in Figures (a)-(d): the emergence of the liquid sheet from the droplet (a), radial expansion (b), formation of liquid fingers at the front of the liquid (c) and droplet detachment (d).}
\label{fig:wedge}
\end{figure}

Using the sheet velocity and thickness, the aerodynamic lifting force (see Fig.~\ref{fig:wedge}), consisting of the suction ($\sim K_l \mu_g V_t $) and lubrication force ($\sim K_u \rho_g V_t^2 H_t $), can be determined. Here, $\mu_g$ and $\rho_g$ are the viscosity and density of the air, respectively. The suction force is caused by the negative pressure difference above the liquid sheet due to Bernoulli principle, while the lubrication force is generated by the air moving underneath the liquid sheet, creating a positive pressure difference that pushes the lamella upward. $K_u$ and $K_l$ are constants. While $K_u \simeq 0.3$ was determined in \cite{Riboux2014} by numerical calculations, $K_l$ can be calculated using the sheet thickness, mean free path of the molecules $\lambda$ in the surrounding air and the wedge angle $\alpha$, which is the angle between the lifted sheet and the surface:

\small
\begin{align}
K_l = -\frac{6}{\tan^2 \alpha}\left( \ln \left(19.2 \frac{\lambda}{H_t}\right) -\ln\left(1+19.2 \frac{\lambda}{H_t} \right) \right)
\label{eq:kl}
\end{align}
\normalsize
According to \cite{Riboux2014}, the wedge angle is equal to $60^\circ$ and should be dependent on the wetting properties of the surface \cite{Riboux2014}. \\

Having determined these parameters, a dimensionless number defined as the splashing ratio $\beta$ is calculated, which indicates the magnitude of the aerodynamic forces needed to overcome the surface tension in order to break up the liquid sheet into smaller droplets:

\begin{align}
\beta &= \left( \frac{K_l \mu_g V_t + K_u \rho_g V_t^2 H_t}{2\sigma} \right)^\frac{1}{2}
\label{eq:zeta}
\end{align}

Comparing Equations \eqref{eq:numte} and \eqref{eq:zeta} to both their own experiments and previous work \cite{Xu2005,Palacios2013,Stevens2014}, Riboux and Gordillo determined find that the value of the splashing ratio should be around $0.14$.\\

To calculate the splashing velocity from the splashing model of Riboux and Gordillo, Equations \eqref{eq:vtfull} and \eqref{eq:htfull} are substituted into Eq.~\eqref{eq:zeta}, for which a quadratic equation for the splashing velocity can be found:

\small
\begin{align}
{\displaystyle
v_{\rm sp} = \frac{ -\frac{\sqrt{3}}{2} K_l \mu_g t_e^{-\frac{1}{2}} + \sqrt{ \frac{3}{4} \left(K_l \mu_g \right)^2 t_e^{-1} + \frac{6\sqrt{3}}{\pi} K_u \rho_g D_0 t_e^\frac{1}{2} \sigma \beta^2}}{ \frac{3\sqrt{3}}{2\pi} K_u \rho_g D_0 t_e^\frac{1}{2} } }
\label{eq:fullmodel}
\end{align}
\normalsize

where $v_{\rm sp}$ is the splashing velocity.\\

To compare the splashing model with our experiments, we first calculate the ejection time for each ethanol-water mixture by substituting Eq.~\eqref{eq:fullmodel} into Eq.~\eqref{eq:numte}. From the obtained ejection time, the splashing velocity can be determined using Eq.~\eqref{eq:fullmodel}. The best fit of the splashing model on our data (green line, Fig.~\ref{fig:vcrit}) was determined by minimising the sum of square residuals using the splashing ratio as a fit parameter. The obtained best fit value is equal to $0.127 \pm 0.004$, which is close to the value 0.14 found by Riboux and Gordillo. While the splashing model has been verified for borosilicate glass substrates before \cite{Stevens2014,Palacios2013}, our data for splashing on stainless steel and parafilm indicate that splashing should be independent of the wetting properties of the impacted surfaces.\\

The wetting properties of the surface should, according to \cite{Riboux2014}, determine the wedge angle $\alpha$.  Riboux and Gordillo however gave no argument why the value of the wedge angle should be $60^\circ$. Therefore, to investigate the time dynamics of the wedge angle during splashing, we measure the wedge angle: in our experiments, we let ethanol droplets impact a stainless steel surface at an impact velocity comparable to the splashing velocity ($v=v_{\rm sp}$). Recording the impact with the high-speed camera gives the time evolution of the liquid sheet expansion, which is depicted in Figures \ref{fig:wedge1} to \ref{fig:wedge4}. We observe that after the moment of the sheet ejection (Fig.~\ref{fig:wedge1}), the liquid sheet starts to radially expand outward (Fig.~\ref{fig:wedge2}). After a certain time, liquid fingers start to form at the edge of the sheet (Fig.~\ref{fig:wedge3}), after which satellite droplets detach from the sheet (Fig.~\ref{fig:wedge4}): splashing occurs.\\

From each recorded frame we extracted the instantaneous wedge angle, which is shown in Fig.~\ref{fig:wedgepoints}. In this graph, the wedge angle seems to increase linearly with time up until the moment of finger formation (Fig.~\ref{fig:wedge3}). Assuming that the wedge angle keeps increasing linearly until droplet detachment, we can fit a simple linear function ($\alpha = A t$) to the data of Fig.~\ref{fig:wedgepoints} and extrapolate the wedge angle to the moment the droplets detach from the liquid sheet. Using this method, we obtain an average wedge angle of $\alpha = 59^\circ \pm 8^\circ$ at the moment of droplet detachment (Fig.~\ref{fig:wedge4}), which is almost identical to the value of the wedge angle postulated by Riboux and Gordillo. These results therefore show that the wedge angle given in \cite{Riboux2014} should be the wedge angle at the moment of droplet detachment, since the angle various continuously in time. The wedge angle at the moment of droplet detachment does not depend on the surface or the fluid, it is simply given by the air viscosity, which explains why the wetting properties are unimportant: the droplet appears to land on an air cushion. It was shown recently that the maximum radius of impacting drops in this high velocity impact regime does not depend on the substrate properties (wettability, roughness) either \cite{Lee2015}, for the same reason.\\

Then, taking a constant wedge angle of $\approx 60^\circ$, we can simplify the splashing model. Most practical situations deal with fluids with a low Ohnesorge number ($Oh \ll 1$) and take place at atmospheric conditions. If we evaluate Eq.~\eqref{eq:numte} for low Ohnesorge numbers, the first term on the left side in the equation dominates, allowing us to write the ejection time as a function of the Weber number $We$ $\left(We= Oh^2 \cdot Re^2 = \frac{\rho R v^2}{\sigma}\right)$ as:

\begin{align}
t_e &\approx We^{-2/3}
\end{align}

Consequently, the lamella thickness and spreading velocity are given by:

\begin{subequations}
\begin{align}
V_t &= \frac{\sqrt{3}}{2}v_{\rm sp} t_e^{-\frac{1}{2}} = \frac{\sqrt{3}}{2} v_{\rm sp} We^{1/3}
\label{eq:vtsimpl}
\end{align}
\begin{align}
H_t&= R \frac{\sqrt{12}}{\pi} t_e^\frac{3}{2} \approx \frac{\sqrt{3}}{\pi} D_0 We^{-1}
\label{eq:htsimpl}
\end{align}
\end{subequations}
where $R=\frac{1}{2}D_0$. \\

A further simplification can be made for the constant $K_l$ [Eq.~\eqref{eq:kl}]. If $H_t$ is calculated with the measured splashing velocity and initial diameter of the drops, an average value on the order of $10^{-5}$ m is obtained. Since $\lambda \sim 10^{-8}$ m , the ratio $\frac{\lambda}{H_t}$ is in the order of $10^{-3}$. As $\frac{\lambda}{H_t}$ is small, the second logarithmic term can be approximated with the first term of the Taylor approximation:

\begin{align}
 \ln \left(1+19.2 \frac{\lambda}{H_t} \right)  \approx 19.2 \frac{\lambda}{H_t} 
\end{align} 

Since $\left| \ln\left(19.2 \frac{\lambda}{H_t}\right) \right| \gg \left| 19.2 \frac{\lambda}{H_t} \right|$ for $\frac{\lambda}{H_t} \sim 10^{-3}$, the above term can be neglected, giving a simplified equation for $K_l$:

\begin{align}
K_l \approx \frac{-6}{\tan^2 (\alpha)}\log\left( 19.2 \frac{\lambda}{H_t} \right) \approx \frac{9.9}{\tan^2 (\alpha)}
\label{eq:Klsimpl}
\end{align}

A final assumption, that was also suggested by Riboux and Gordillo, is that the lubrication force dominates over the suction force under atmospheric conditions \cite{Riboux2014}, implying that Eq.~\eqref{eq:zeta} can be simplified to:
\begin{align} 
\beta &\approx \left( \frac{K_l \mu_g V_t }{2\sigma} \right)^\frac{1}{2}
\label{eq:zeta2}
\end{align} 

Finally, by substituting the simplified terms of $V_t$ [Eq.~\eqref{eq:vtsimpl}] and $K_l$ [Eq.~\eqref{eq:Klsimpl}], $\beta$ can be rewritten as a function of the splashing velocity $v_{\rm sp}$, initial diameter $D_0$, density $\rho$, surface tension $\sigma$, wedge angle $\alpha$ and the viscosity of the air $\mu_g$:

\begin{align} 
\beta \approx 2.07\frac{\sqrt{\mu_g}}{\tan \alpha} \left(\rho D_0 \right)^{1/6} \frac{v_{\rm sp}^{5/6}}{\sigma^{2/3}}
\label{eq:approxzeta}
\end{align} 

Thus, it is possible to significantly simplify Riboux and Gordillo's splashing model for low Ohnesorge number fluids and assuming atmospheric conditions, where the splashing ratio is only dependent on the fluid parameters and the viscosity of the air, as the wedge angle seems to be identical for all smooth surfaces.\\

In order to compare the simplification with the experimental data, Eq.~\eqref{eq:approxzeta} can be rewritten into a simple analytical expression for the splashing velocity:
\begin{align}
v_{\rm sp} = \left( \frac{\tan \alpha \hspace{0.1cm} \sigma^\frac{2}{3} \beta}{2.07 \hspace{0.1cm} \sqrt{\mu_g} \left( \rho D_0 \right)^\frac{1}{6}} \right)^\frac{6}{5}
\label{eq:critsimpl}
\end{align} 

Then, the splashing velocity only depends on the density, surface tension and initial diameter of the droplet, the viscosity of the air, the wedge angle and the splashing ratio. The simplification (red dashed line) is plotted together with the experimental data and the full splashing model in Fig.~\ref{fig:vcritc}. Again, the splashing ratio was used as fitting parameter, where the best fit value of $\beta$ is equal to $0.127 \pm 0.003$, identical to the splashing ratio of the full splashing model. The simplification also gives an equally good prediction of the measured splashing velocity. Furthermore, the predicted values of the full splashing model and the simplification are similar, with the largest relative error of around $17.7 \hspace{0.08cm} \%$ \footnote{Here, the relative error is defined as $\frac{v_{simpl} - v_{full}}{v_{full}}\cdot 100\%$, where $v_{simpl}$ and $v_{full}$ are the splashing velocity predicted by the simplification and full splashing model, respectively}. Therefore, this result shows that the splashing velocity can be predicted easier compared to the full splashing model.\\

To summarise, we systematically investigated the influence of the surface tension of the droplet and wetting properties of the surface on the splashing velocity of droplets impacting a smooth surface. We showed that the wetting properties do not influence the splashing velocity. Second, we compared experimental with a preexisting splashing model and showed that the model can be applied to both hydrophilic as hydrophobic surfaces. By measuring the wedge angle, we confirmed that the wedge angle used in the splashing model is equal $60^\circ$ at the moment of droplet detachment from the liquid sheet. Finally, we proposed a simplification on the splashing model based on low Ohnesorge numbers and atmospheric conditions and obtained a simple equation for the splashing velocity, which both agree very well with the experimental data. \\

\section*{Acknowledgements}
We would like to thank Detlef Lohse, who pointed us towards the work of Guillaume Riboux and Jos\'e Manuel Gordillo.
\newpage

\end{document}